\documentclass[10pt,twocolumn,final]{asme2ej}

\usepackage{epsfig} 
\usepackage{graphicx}
\usepackage{amsmath}
\usepackage{upgreek}
\usepackage{threeparttable}
\usepackage{amssymb}

\usepackage{fancyhdr}
\usepackage{helvet}
\usepackage{hyperref}   
\hypersetup{
	colorlinks=true,
	linkcolor=blue,
	citecolor=blue,
	urlcolor=blue,
}

\title{Numerical simulation of thermal conductivity of stainless steel and Al-12Si powders for additive manufacturing}

\author{Vladimir~Ankudinov \thanks{Corresponding author}
    \affiliation{
    Udmurt State University\\
    Institute of Mathematics, Information Technologies\\and Physics\\
     Izhevsk, 426034\\
     Russia\\
     \\
     Vereshchagin Institute for High Pressure Physics\\
     Russian Academy of Sciences\\
     Moscow (Troitsk), 108840\\
     Russia \\
    Email: vladimir@ankudinov.org
    }	
}

\author{Georgii A.~Gordeev
    \affiliation{
    Udmurt State University\\
    Institute of Mathematics, Information Technologies\\and Physics\\
     Izhevsk, 426034\\
      Russia\\
   Email: gordeevgeorgii@gmail.com}
}

\author{Evgeniy V.~Kharanzhevskiy
    \affiliation{
    Udmurt State University \\
    Institute of Mathematics, Information Technologies\\and Physics\\
     Izhevsk, 426034\\
     Russia\\
   Email: eh@udsu.ru}
}

\author{Mikhail D.~Krivilyov
    \affiliation{
    Udmurt State University \\
    Institute of Mathematics, Information Technologies\\and Physics\\
     Izhevsk, 426034 \\
     Russia\\
     \\
   Udmurt Federal Research Center\\
    Ural Branch of the Russian Academy of Science\\
    Izhevsk, 426067\\
    Russia
    Email: mk@udsu.ru
    }	
}

\begin{document}

\maketitle    

\begin{abstract}
{\it 
A three-dimensional model of a partially melted powder bed with particles stochastically distributed in size and space coordinates has been developed. Numerical simulation of temperature distributions in stainless steel AISI 316L and Al-12Si powders in vacuum, air and argon has been performed to analyze unsteady heat transfer in a porous medium. The numerical model demonstrates a large effect of  heat transfer through the gas phase in case of powders with low thermal conductivities like stainless steels. At the porosity level of 65\% and above, the mechanism of heat transfer drastically changes and a linear dependence of thermal conductivity on porosity frequently used in literature becomes incorrect. The effects of the consolidation coefficient and size distribution on effective heat transfer in powders are discussed. The obtained dependencies of the effective thermal conductivity on porosity and the consolidation coefficient could be used in additive manufacturing applications. 

}
\end{abstract}


\section{Introduction}

\begin{figure}
\begin{center}
\includegraphics[width=0.49\columnwidth]{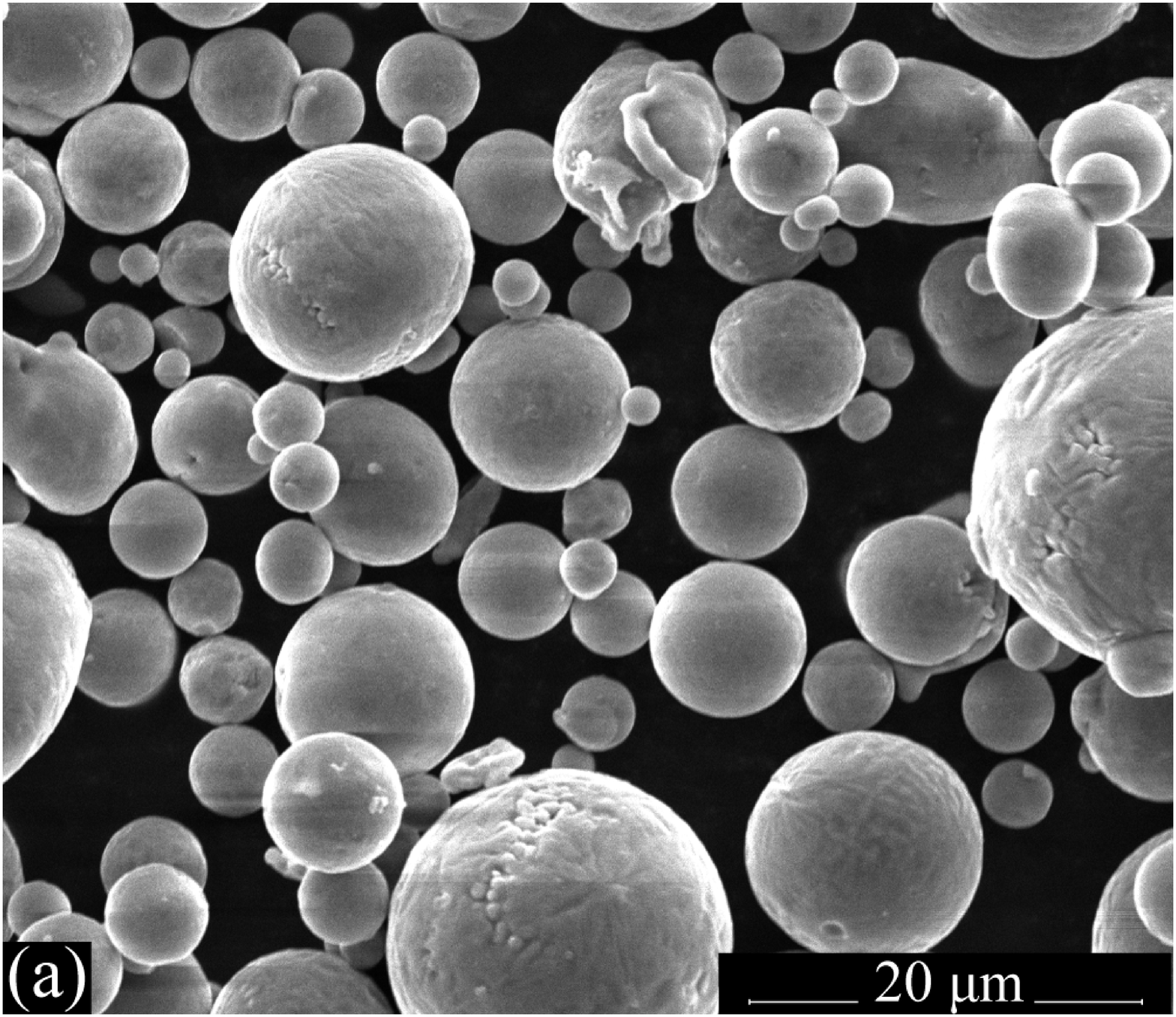}\hspace{0.005\columnwidth}
\includegraphics[width=0.49\columnwidth]{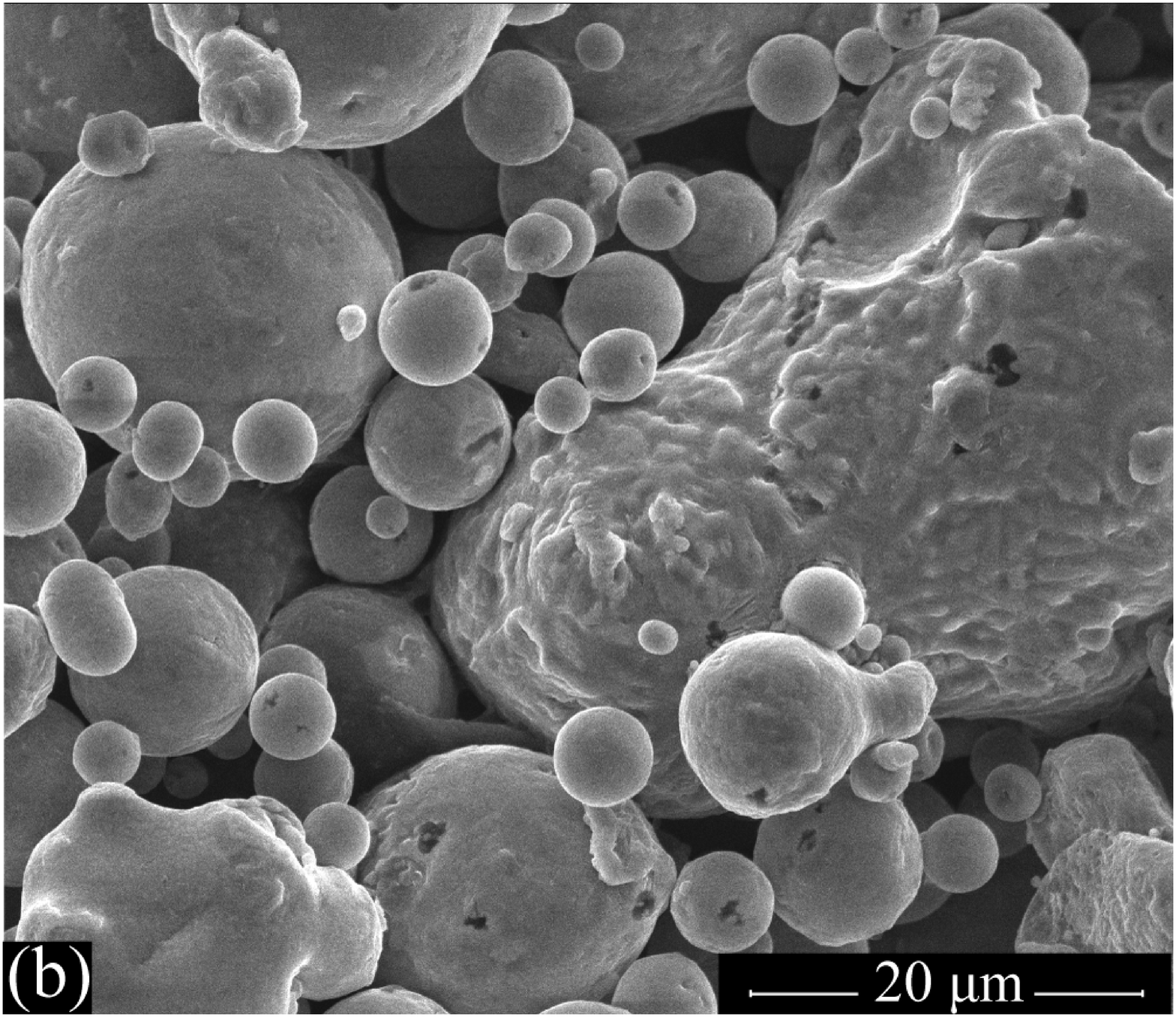}
\end{center}
\caption{\label{fig1}
(a) SEM image of raw commercial Al12SiMg SLM powder, grade CL30AL~\cite{cl} studied in the present work; (b) the same powder exposed for 1 hour heat treatment at $T=623$~K. The necks between powder particles due to solid-state consolidation are formed and coagulation of the particle's conglomerates started. The SEM voltage is 10kV at magnification of 5000X.}
\end{figure}

Transport of heat through a porous medium is of great interest in  relation with its morphological  properties since in many technological processes effective thermal properties control the dynamics of heat conduction, melting and solidification~\cite{king15, yap15}. The important technical applications where the metallic powders are used include powder metallurgy, selective laser sintering (SLS) and  selective laser melting (SLM)~\cite{shishkovskiy16}. Usually, the structure of a porous media is complex owing to a wide distribution of particles in size and irregular morphology of pores in a metal matrix~\cite{hsu94,kaviany,likov67}. Due to the fact that the commonly utilized in SLM  powders are spheroidized (see scanning electron microscopy (SEM) image of raw commercial Al-based alloy powder in Fig.~\ref{fig1}(a)), the simplest particle's approximation by spheres can be used~\cite{shishkovskiy16,kaviany}. This approximation  substantially simplifies the results of thermal analysis for real industrial powders~\cite{kaviany}. Thus, the detailed prediction of the effective thermal conductivity of powder media requires consideration of the shape and size distribution, conductivity of each particle and gas atmosphere, and heat exchange between the metal and gas phases~\cite{zehnder70}. In the most cases, for an anisotropic porous medium saturated by gas or liquid the thermal conductivity has significantly  lower  values  than for the same non-porous material~\cite{chen2000, likov67}.
The thermal conductivity of dry powders could differ significantly for Van-Der-Waals force bonded, ionic-bonded or covalent-bonded materials~\cite{kaviany,Chan1973,Stewart1973}. This difference comes from the contact resistance effect between particles which plays a significant role and is ruled mainly by three factors: (i) elastic deformation of the particles caused by the external compressive force, (ii)  chemical composition of the particle’s surface, and (iii) Kapitza resistance. The thermal contact resistance is successfully studied for the case of particles with low-area contacts. However, in the case of SLM processing the formation of improved thermal contacts between particle's  surfaces due to solid-phase, liquid-phase sintering and pre-melting leads to the reduction of these effects.

Due to surface pre-melting of particles during laser treatment, the neighboring particles start to form necks between them which significantly improves heat exchange in a porous medium~\cite{Gibson2013, shishkovskiy16}, see Fig.~\ref{fig1}(b).  The improved thermal contact between particles or the effect of dynamical consolidation under conditions of laser treatment may significantly affect the effective heat transport in the processed powder bed. To precisely control processing parameters, one can account for the above-mentioned effect by introduction of a consolidation coefficient which influences the effective thermal conductivity~$k_{eff}$~\cite{hadley86}.
Then simulation of SLM processes may be carried out in the approximation of a continuous media~\cite{Gordeev2017,Gordeev2011,hodge14,huang16,dong09,Romano2015,Gordeev2020}  where the effective heat conductivity coefficient is treated as a state functions.

The existing approaches in modelling of heat transfer in powder beds can be reduced to two major cases: \\
(i)  discrete particles modelling where the powder bed is  described as a domain filled with the separate particles \cite{alkahari12,kaviany,holzbecher12,wang16,yang16,Moser2016,Lee2018};  \\
(ii) elementary cell approach where heat transfer in one particle and its elementary volume is considered as periodically repeated in the whole volume \cite{kaviany,ofuchi64,kandula11,chen2000, chen15, yang10, asakuma12, slavin02, hsu94,gusarov10,gusarov09}.
\\
To establish a given porous medium's properties using the discrete particles approach (i) one have to measure averaged sample's properties. If the sample is too small, parameters such as heat conduction coefficients and other mass-related parameters tend to oscillate. With the increasing of the sample size, the oscillations begin to dampen out. The sample size related to the consistent averaged properties and designated as representative volume element (RVE) when its elastic constants, porosity and thermal characteristics are the same as for the composite or powder medium \cite{Hill1963,kaviany,shishkovskiy16}.
Calculation of the effective heat conductivity in the powder with spherical particles can be done by combination of the classic packed bed approach~\cite{hsu94,zehnder70} and the stochastic random filling models (see~\cite{kaviany} and references therein). This approach and its comparison to the existed analytical models is presented in this article.

The objective of this paper is to study the effect of powder bed morphology, porosity and particle consolidation on the effective thermal conductivity of commercial Al-12Si and stainless steel SLM powders during the initial and intermediate stages of heating or melting/sintering. The numerical study of heat transfer is carried out for randomly generated powder beds filled with spherical particles of different sizes at different consolidation levels in vacuum, air and argon. We shall consider this process as solid-state consolidation, though in general the dynamics of powder compaction involves also solid-liquid phase transitions and motion dynamics of phase interfaces~\cite{Khairallah2016,king15,yap15,Gordeev2020}.


\begin{table*}[t]
\caption{Technical specification of Al-12Si and stainless steel 316L powders and thermophysical parameters used in the present paper.}
\begin{center}
\begin{threeparttable}
\begin{tabular} {p{45mm}p{50cm}p{50mm}}
\hline
\multicolumn{1}{l|}{Parameter}                                 & \multicolumn{2}{c}{Value}                                                   \\ \hline
\multicolumn{1}{l|}{}                                              & \multicolumn{1}{l|}{Al-12Si}                            & Stainless steel   \\ \cline{2-3}
\multicolumn{1}{l|}{Commercial name}                               & \multicolumn{1}{l|}{CL30AL}                             &      CL20ES, AISI 316L             \\
\multicolumn{1}{l|}{Material}                                      & \multicolumn{1}{l|}{Al-based, 10.5..13.5 wt.\% Si}                & Fe, $0.03$ wt.\% C, $16..18$ wt.\% Cr, \\
\multicolumn{1}{l|}{}                                              & \multicolumn{1}{l|}{0.05 wt.\% Mn, 0.25 wt.\% Fe~\cite{cl}}   & $10..14$ wt.\% Ni, $2..3$ wt.\% Mo~\cite{cl}  \\
\multicolumn{1}{l|}{Shape of particles}                                         & \multicolumn{2}{c}{Quasi-spherical or\ spherical}                                  \\
\multicolumn{1}{l|}{Mean\,/\,min\,/ max diameter, $\langle d \rangle$, $\upmu$m}               & \multicolumn{1}{l|}{4.9 / 1.3 / 18.6\tnote{a}} &  37.5 / 15.0 / 50.0  \cite{Kabbur2017,Spierings2011,Choi2016} \\
\multicolumn{1}{l|}{Bulk thermal conductivity, $k^\textrm{metal}$, W/(m$\cdot$K)} & \multicolumn{1}{l|}{155 \cite{cl,tang}}           & 15 \\
\multicolumn{1}{l|}{Bulk heat capacity, $C^\textrm{metal}_p$, J/(K$\cdot$kg)}  & \multicolumn{1}{l|}{947 \cite{crc,sekulic}}                   & 500 \cite{cl}\\
\multicolumn{1}{l|}{Bulk density, $\rho^\textrm{metal}$, kg/m$^3$}           & \multicolumn{1}{l|}{2640 \cite{nikan,crc}}                         & 8000 \cite{cl}      \\
\multicolumn{1}{l|}{Melting temperature, $T_m$, K}           & \multicolumn{1}{l|}{854 \cite{crc,sekulic}}                         &1713 \cite{cl,crc}        \\ \hline
\multicolumn{1}{l|}{}                                              & \multicolumn{1}{l|}{Air}                            & Argon  \\ \cline{2-3}
\multicolumn{1}{l|}{Gas thermal conductivity, $k^\textrm{gas}$, W/(m$\cdot$K)} & \multicolumn{1}{l|}{2.42$\times 10^{-2}$ \cite{crc,nist}}           & 1.78$\times 10^{-2}$ \cite{crc,nist} \\
\multicolumn{1}{l|}{Gas heat capacity, $C^\textrm{gas}_p$, J/(K$\cdot$kg)}      & \multicolumn{1}{l|}{1005 \cite{crc,nist}}                   & 519 \cite{crc,nist}  \\
\multicolumn{1}{l|}{Gas density, $\rho^\textrm{gas}$, kg/m$^3$ }         & \multicolumn{1}{l|}{1.29 \cite{crc,nist}}                         & 1.63 \cite{crc,nist}       \\ \hline
\end{tabular}
   \begin{tablenotes}
      \item[a]{Obtained in the present work}
    \end{tablenotes}
\label{powderpars}
\end{threeparttable}
\end{center}
\end{table*}

\begin{figure}
\begin{center}
\includegraphics[width=0.92\columnwidth]{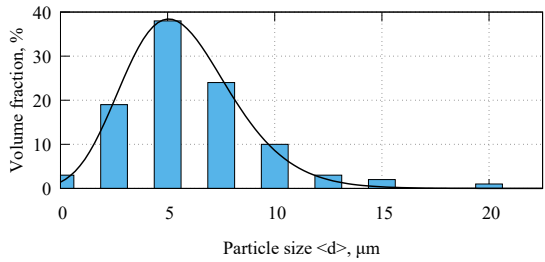}
\end{center}
\caption{\label{figsizedistr}
Particle size distribution of raw commercial Al-12Si SLM powder CL30AL~\cite{cl} evaluated on the basis of SEM images Fig.~\ref{fig1}(a). The distribution is approximated by the lognormal probability density function (solid line) with the parameters of $\sigma=0.25$, $\mu = 0$, shifted and normalized.}
\end{figure}

\section{Model of effective heat transfer in randomly allocated  powder beds}
In this section, a mathematical model of heat transfer in the partially-melted (consolidated) metallic powder beds is formulated. The powder bed is generated using random stochastic deposition of particles in the representative volume element (RVE) with up to $10^3$ metallic particles. Using direct simulations of heat transfer in such layer, one can find the dynamics of heat propagation in a metallic powder processed in gas atmosphere or in vacuum for the given porosity~$\varepsilon$, particle size distribution and particle contact spot diameter~$d_C$. The model allows one to choose the consolidation coefficient of particles and the bed porosity as input parameters during generation of powder filling.

\subsection{Analysis of local thermal equilibrium}
\label{sec:analysis}

For macroscopic simulations of transport phenomena in discontinuous or two-phase systems the microscopic governing equations are averaged using the coarse-graining procedure~\cite{Samaey}. Additionally, the effective transport characteristics of powder media are described in the approximation of continuous media~\cite{kaviany,likov67}. These characteristics like the effective thermal conductivity depend on temperature, porosity and other powder parameters and may be implemented as state functions during modeling of powder shrinkage and melting~\cite{dong09, Gordeev2017, Gordeev2011, hodge14, huang16, Romano2015, Gordeev2020}. To that end, three important aspects are to be considered as follows for proper space and time discretization in numerical simulation of heat transfer in porous media.

(i) \textbf{Local thermal equilibrium.} The classic approach~\cite{kaviany, virto09} for modeling is based on justification of the condition of local thermal equilibrium between gas and powder particles. The general condition for local thermal equilibrium in a discrete volume element can be formulated as~\cite{kaviany}:
\begin{equation}
\frac{1}{V^\textrm{gas}} \int_{V^\textrm{gas}} T^\textrm{gas} dV = \frac{1}{V^\textrm{metal}} \int_{V^\textrm{metal}} T^\textrm{metal} dV,
\label{eq:termo-loc-equi}
\end{equation}
where $V=V^\textrm{gas}+V^\textrm{metal}$ is the total volume of a discrete element, the superscripts ``gas'' and ``metal'' belong to the designate phases, and $T$ is local temperature.

(ii) \textbf{Representativity of discrete volumes.} First, the effective thermal conductivity of a discrete volume should be independent on the size of this volume. This criterion allows one to find a size of the representative volume element (RVE) and the upper  limit of the mesh size for discretization of the powder bed. Typically, RVE contains no less than $5^3..7^3$ particles~\cite{kaviany}. Figure~\ref{fig1} provides the SEM images of Al-12Si powder at initial state and after solid-phase sintering while the size distribution and mean diameter are provided in Fig.~\ref{figsizedistr}. The specifications of two SLM powders are given in Table~\ref{powderpars}. The particle size has an average of $\langle d \rangle = 4.9$ $\upmu$m. Therefore, one can estimate that a linear size of RVE is about $L_\textrm{RVE} = 25..35 $~$\upmu$m. Second, representativity means continuity of the thermal field. Continuity is  satisfied in the finite element method, since regularity of the temperature field in each finite element is automatically provided owing to linearization of the non-stationary second-order  conduction equation in the case of zero heat source~\cite{Segerlind1984}.

(iii) \textbf{Consistency of discrete volumes with macroscopic characteristic lengths.}
Let one consider the local temperature difference between gaseous and metallic phases to be $\Delta T^\textrm{gas/metal}$. The maximum attainable temperature difference in the RVE is $\Delta T_\textrm{RVE} $.  RVE fulfills the upper limit for discretization on the macroscopic scale~$L$ if the general temperature difference $\Delta T_{L}$ is much higher than one in the RVE
\begin{equation}
\Delta T^\textrm{gas/metal} <  \Delta T_\textrm{RVE} \ll \Delta T_{L}.
\label{eq:rve-condition}
\end{equation}
The typical temperature difference attained during SLM processing is as follows~\cite{Gordeev2017,Gordeev2020}:
\begin{align}
\Delta T^\textrm{gas/metal} & \sim 20..50\, K,\\
\Delta T_\textrm{RVE} &\sim 150\,  K,\\
\Delta T_{L} &\sim 500..600\, K.
\end{align}
Therefore, to consider media in the RVE in the continuous approximation the temperature criterion Eq.~(\ref{eq:rve-condition}) satisfies if $L>200$~$\upmu$m.

(iv) \textbf{Time discretization.} For time discretization, the condition Eq.~(\ref{eq:termo-loc-equi}) of local thermal equilibrium~\cite{carbonell84, kaviany} has to be applied to a single discrete element. This condition formulated for heat propagation in connection to a time step $\delta t$ in the discrete element with a size of $ \ell $ may be formulated as
\begin{equation}
\frac{\varepsilon (C_p \rho )^\textrm{gas}\, \ell^2}{\delta t} \left( \frac{1}{k^\textrm{gas}} + \frac{1} {k^\textrm{metal}}  \right) \ll 1,
\label{eq:termo-loc-equi-c0}
\end{equation}
\begin{equation}
\frac{(1-\varepsilon)  (C_p \rho )^\textrm{metal}\, \ell^2}{\delta t} \left( \frac{1}{k^\textrm{gas}} + \frac{1} {k^\textrm{metal}}  \right) \ll 1.
\label{eq:termo-loc-equi-c1}
\end{equation}
This time and length scales must be satisfied for the system at local thermal equilibrium, requiring the heat fluxes to be small enough to calculate the thermal conduction equation. If the size of discrete elements is $\ell=1~\upmu$m, then the time step $\delta t$ in calculations should be of the order of $\delta t \sim 10^{- 8}$~s or lower to meet the conditions Eqs.~(\ref{eq:termo-loc-equi-c0})--(\ref{eq:termo-loc-equi-c1}) for Al-12Si SLM powder. Finally, the particles of $\langle d \rangle = 5 $~$\upmu$m must be discretized by  at least 5 finite elements. In the present work, we chose the size of the adaptive mesh elements to be~$\ell\sim 0.05..1$~$\upmu$m and the time step of about $\delta t \sim 10^{- 8}$~s to fulfill Eqs.~(\ref{eq:termo-loc-equi-c0})--(\ref{eq:termo-loc-equi-c1}).

\subsection{Powder bed generation}
\label{gen}
\begin{figure}[h]
\begin{center}
\includegraphics[width=0.6\columnwidth]{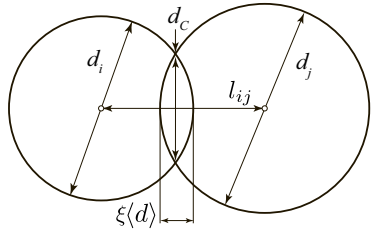}
\caption{\label{f1}
Schematic view of the contact between two consolidated particles in partially-consolidated powders. The diameters of two particles are $d_i$ and $d_j$ correspondingly, $d_C$ is the contact spot diameter, $\langle d \rangle$ is the mean diameter of two particles, $l_{ij}$ is the distance between the sphere's centers and $\xi$ is the overlapping coefficient according to Eq.~(\ref{eq:xixi}).}
\end{center}
\end{figure}

\begin{figure}
\begin{center}
\includegraphics[width=0.49\columnwidth]{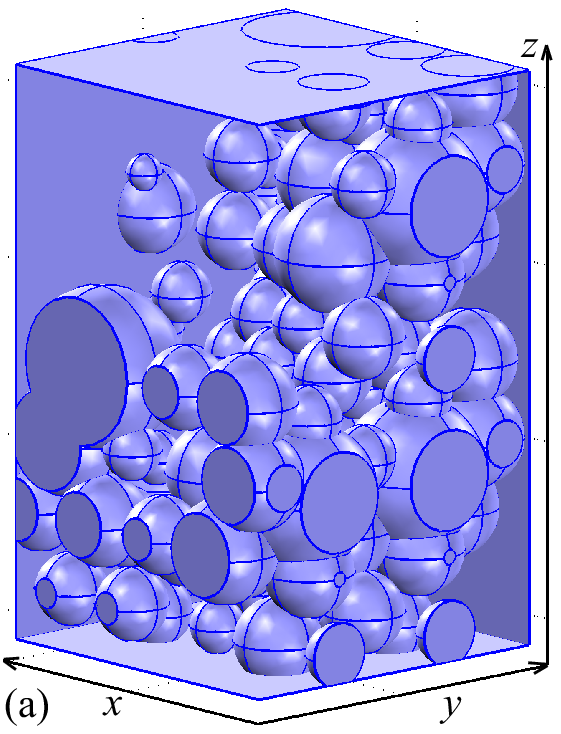}\hspace{0.05\columnwidth}
\includegraphics[width=0.42\columnwidth]{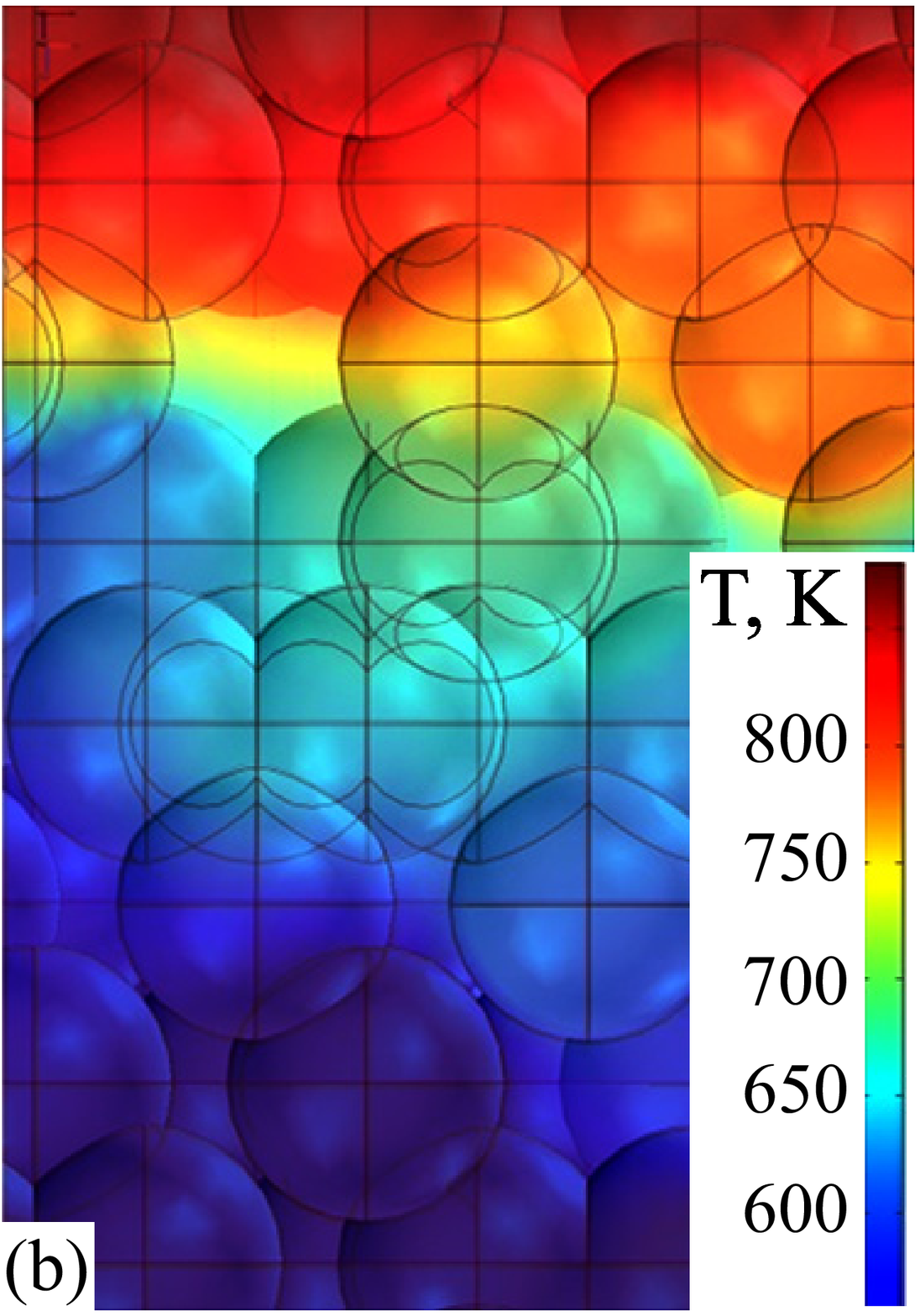}
\caption{\label{heatfield}
(a) Generated particle bed with the log-normal distribution of  particle sizes (see Fig.~\ref{figsizedistr}) at $\langle d \rangle=4.9$~$\upmu$m, $\varepsilon=0.72$.
(b) Projection of the instantaneous thermal field $T$ in the Al-12Si  powder bed  calculated for SLM in air with a fixed particle size of $\langle d \rangle=4.9$~$\upmu$m,  size deviation of $\sigma=0$, contact diameter of $d_C = 3$~$\upmu$m, and porosity of $\varepsilon=0.65$. The physical time is $t=3.05 \times 10^{-5}$~s. The evaluated effective heat conductivity is obtained by Eq.~(\ref{sol}) and it equals to $k_{eff}=16.2$~$\textrm{W}/(\textrm{m} \cdot \textrm{K})$. }
\end{center}
\end{figure}

The powder bed was generated by placing randomly the spherical particles in nodes of a hexagonal closed-pack (HCP) lattice. We assumed the realistic log-normal size distribution~\cite{king15,gusarov10} with the parameters of~$\mu=0$, $\sigma=0.25$, see Fig.~\ref{figsizedistr}. This size distribution parameters correspond to the morphology of commercial SLM CL30AL powder. The diameter~$d_C$ of the contact spot and thus dimensionless overlapping coefficient~$\xi$ between particles was controlled by proper displacement of each particle
\begin{align}
\label{eq:xixi}
\xi  =\frac{\langle d \rangle - l_{ij}}{\langle d \rangle }, \quad \quad \langle d \rangle =\frac{d_i+d_j}{2},
\end{align}
where~$d_i, d_j$ are the sizes of the particles $i$ and $j$, $\langle d \rangle$ is the mean diameter of two particles, $l_{ij}$ is the distance between their centers, see Fig.~\ref{f1}. Then the contact spot diameter~$d_C$ could be found from
\begin{align}
16 d_C^2 l_{ij}^2 = \left( (d_i+d_j)^2 - 4 l_{ij}^2 \right) \times \left( -(d_i-d_j)^2 + 4 l_{ij}^2 \right).
\end{align}
The specified porosity was achieved by removal of random particles from the bed till the target average porosity $\varepsilon=V^\textrm{gas}/V$ is  attained. Here $V$ is the domain volume and $V^\textrm{gas}$ is the volume of the gaseous phase only.
Due to the HCP packing limit, generation of powder beds at low porosities $\varepsilon < 0.26$  was performed by random allocation of the gas voids between particles which reproduces a powder briquette or a partially consolidated high-density powder bed.
The porosity obtained for dry powder beds experimentally varies between $\varepsilon=0.41$ and $0.46$ for loose-packed stainless steel \cite{Choi2016,Jafari2018}.
Once the particles are deposited, their positions and radii are used as input parameters for generation of the finite element mesh and implementation of the heat transfer model. A heat source was allocated at the top boundary of the computational domain with dissected particles (see Fig.~\ref{heatfield}).

\subsection{Governing equations and computational domain}
The transient problem of heat conduction in a two-phase ``gas-metal'' domain was formulated. The effect of heat transfer by radiation between particles was neglected due to its small contribution to the overall energy balance. The convective mechanism of heat exchange through the gaseous phase was also eliminated for the specific case at the considered time scales $\tau=10^{-8}-10^{-5}$\,s because viscosity of the gas phase significantly reduces convection in small pores~\cite{kaviany,likov67}. More details are provided below in Sec.~\ref{conv}.
Accounting for nice contact between the pre-melted particles with formed necks we neglect the surface effects and contact thermal resistance since its influence is drastically reduced with compaction evolvement in SLM-processing.
Heat conduction in the gaseous and metallic phases correspondingly is described by the following equations
\begin{align}
\label{eq:heat-first}
  C_p^\textrm{gas} \rho^\textrm{gas} \frac{\partial T}{\partial t} & = k^\textrm{gas} \nabla^2 T, \\
C_p^\textrm{metal} \rho^\textrm{metal} \frac{\partial T}{\partial t}& = k^\textrm{metal} \nabla^2 T,
\end{align}
where $C_p$ is the specific heat capacity, $\rho$ is the density, $k$ is the  thermal conductivity of each phase (see correspondent values in Table~\ref{powderpars}). The authors pay attention that the thermal conductivities~$k$ of the phases were selected as constants and are placed aside the $\nabla^2$--operator. In the present paper, this is justified by the fact that the effective thermal conductivity is analyzed as a function of geometric characteristics of the powder bed. Thus, the dependencies of~$k$, $C_p$ and~$\rho$ on temperature are intentionally excluded from analysis and its values is selected for $T_0=293$~K (see Table~\ref{powderpars}). Obtained effective characteristics could be utilized as a base for the modeling of the powder media in the continuous medium approximation when the effective media conductivity is separated in two components:  the effective structural conductivity component  and thermal dependent component resulting $k_\textrm{media}=k_{eff}(\varepsilon,d_C,\langle d \rangle)\cdot k^\textrm{metal}(T)/k^\textrm{metal}(T_0)$ \cite{Gordeev2017,Gordeev2011,Gordeev2020}. This approach allows one to simply describe the complex averaged medium's properties with many dependent parameters at low cost.

The temperature at the boundary $z=0$ was selected as the melting temperature $T_m$.
So the  initial and boundary conditions of the heat conduction problem on the particle's surfaces and domain's walls are
\begin{align}
\label{eq:heat-first-boundary}
T |_{t=0}= T&_{amb},\quad  T |_{z=0}=T_m, \nonumber \\  \mathbf {n} \cdot k  \mathbf{\nabla} T = 0 \,\,\, &\textrm{~at outer boundaries.}
\end{align}
At the interfaces between gas and metal, continuity of the heat fluxes and temperature was defined
\begin{align}
k^\textrm{gas} \, \mathbf{\nabla} T^\textrm{gas} &= k^\textrm{metal} \, \mathbf{\nabla} T^\textrm{metal}, \nonumber \\
T^\textrm{gas} = T^\textrm{metal}\,\,\,& \textrm{~at inner boundaries.}
\label{eq:heat-boundary}
\end{align}
Ambient temperature $T_{amb}$ was equal $0.7 \times T_m$ as far as this value corresponds to the start temperature of the process of solid state consolidation according to~\cite{Meier2017,nikiforov72}.

\subsection{Numerical implementation}
The direct problem Eqs.~(\ref{eq:heat-first})--(\ref{eq:heat-boundary}) of heat transfer in a two-phase powder bed was solved using the finite element method (FEM). The size of mesh elements and time step were selected according to the conditions Eq.~(\ref{eq:termo-loc-equi-c0}) and Eq.~(\ref{eq:termo-loc-equi-c1}) of local thermal equilibrium in RVE. The explicit integration time scheme was used with the backward-differential formula of the 2$^{nd}$ order. This provided sufficient accuracy and smoothness of the time-dependent solution. We generated the spatial mesh consisted up to 10$^6$ tetrahedral Lagrange finite elements with the linear $C^1$ interpolation functions. This mesh provides up to~$3\times10^5$ degrees of freedom. The powder bed generation was first conducted in the MATLAB software and then the subdomain classes which describes particles positions and sizes were invoked in the COMSOL Multiphysics software \cite{comsol2}.
We performed up to 5 simulations with random particle configurations for each set of parameters: porosity $\varepsilon$, contact spot diameter $d_C$,  mean particle diameter $\langle d \rangle$  and size distributions.
The periodic boundary conditions on elongates sides of the computational domain were tested but no significant effect on the temperature field was found if the RVE of a sufficient size is constructed (see Sec.~\ref{sec:analysis}). Different types of the surface layer on the heated top boundary were tested. The difference in the efficient heat conductivity of the powder bed was found negligible. The specific calculations to clarify the role of the dispersion and size distribution of particles were also performed. Finally, the inverse heat transfer problem and its fitting were carried out in MATLAB.

\subsection{Evaluation of the effective thermal conductivity}

The inverse problem of heat conduction~\cite{carslaw84, bird07} was solved using the analytical solution with the complementary error function erfc in order to determine the effective thermal diffusivity $\alpha_{eff}$ as follows
\begin{align}
\label{sol}
\frac{\partial T}{\partial t} = \alpha_{eff} \nabla^2 T, \quad \textrm{which leads to the solution} \\T(z,t)=T_m\!-(T_m\!- T_{amb})\,\textrm{erfc}\!\left( \frac{z}{2\sqrt{ \alpha_{ef\!f}\, t}}   \right)\!\!.
\end{align}
The one-dimensional problem is solved in the semi-infinite approximation admitting that the numerical solution is received in a relatively long in the $z$-direction domain. The three-dimensional formulation of Eqs.~(\ref{eq:heat-first})--(\ref{eq:heat-boundary}) considers the planar heat wave propagation, so temperature was averaged in $xy$-planes leading to the one-dimensional temperature distribution $T^\textrm{FEM}(z,t)$ (see the two-dimensional projection of the $T$-field in Fig.~\ref{heatfield}(b)). Then minimization of  the functional $F(z,t) = \int (T^\textrm{FEM}(z,t)\!\!-T(z,t)) dV$ was carried out.
In the performed simulations, a number of temperature profiles calculated by FEM at fixed times in the gas-metal domain was analyzed.
The calculated effective thermal diffusivity $ \alpha_ {eff} $ were minimized running through the different time frames for each $z$. By this, one can reduce the influence of the boundary particles and their contribution to the $T(z,t)$-distribution and relaxation of the heat wave.
At the end, $\alpha_{eff}$ was averaged for the different initial particle distributions at a given value of $\varepsilon$ and the  effective thermal conductivity of the metallic powder is obtained as
\begin{equation}
k_{eff}(\varepsilon)= \alpha_{eff} \left(C_{p}^\textrm{gas}  \rho^\textrm{gas} \, \varepsilon + C_{p}^\textrm{metal}   \rho^\textrm{metal} \, (1 -  \varepsilon) \right).
\end{equation}

\section{Results}

\subsection{Role of radiative and convective heat transfer}
\label{conv}
\begin{figure}
  \centering
  \includegraphics[width=0.49\columnwidth]{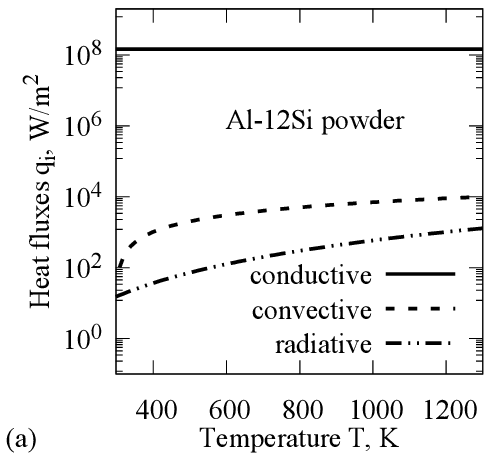}
  \includegraphics[width=0.49\columnwidth]{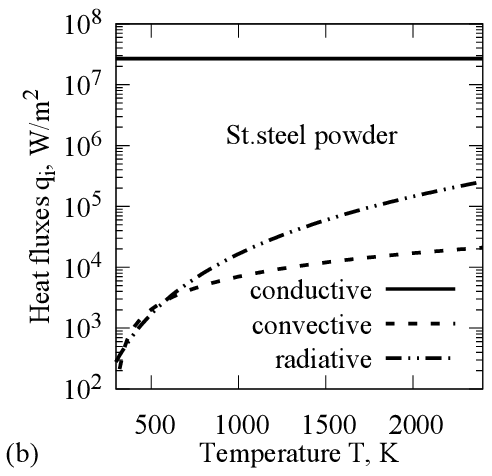}
  \caption{Different contribution in the total heat flux calculated for (a) Al-12Si powder and (b) stainless steel CL20ES powder beds with the porosity $\varepsilon=0.55$. The conductive mechanism Eq.~(\ref{eq:conductive_flux}) is assumed for solid-to-solid conductance. The convective flux Eq.~(\ref{eq:convective_flux}) controls heat transfer between solid particles and gas and dominates in capillary flow inside gas pores. The radiative transfer Eq.~(\ref{eq:radiative_flux}) proceeds between the particles.}\label{fig:heat_fluxes}
\end{figure}

Heat transfer comprises the diffusive, convective and radiative contributions. The effect of radiation on heating of metal powder has been estimated theoretically. According to the Stefan-Boltzmann law, the total emissive power $q_{\text{radiative}}$ of a surface is
\begin{equation}\label{eq:radiative_flux}
  q_{\text{radiative}} = \epsilon \, \sigma_{\text{SB}} \, (T_{\text{surf 1}}^4 - T_{\text{surf 2}}^4),
\end{equation}
where $\epsilon$ is the emissivity ($\epsilon = 0.3$ for Al-12Si and  $\epsilon = 0.7$ for stainless steel), $\sigma_{\text{SB}} = 5.67\times10^{-8}$ \mbox{W m$^{-2}$ K$^{-4}$} is the Stefan-Boltzmann constant, $T_{\text{surf 1,2}}$ are the surface absolute temperatures of two neighbour particles in a powder bed. The heat flux given by  conduction between the contacting particles is defined by Fourier's law as
\begin{equation}\label{eq:conductive_flux}
  q_{\text{conductive}} = k_{eff} G,
\end{equation}
where~$G$ is an estimate of the temperature gradient in the powder bed with the effective conductivity~$k_{eff}$. The conductive heat flux between the particles and gaseous atmosphere follows from a balance of heat exchange between the solid and viscous medium
\begin{equation}\label{eq:convective_flux}
  q_{\text{convective}} = h \Delta T,
\end{equation}
where~$h$ is the surface heat transfer coefficient and~$\Delta T$ is the temperature difference between the particle surface and gas  temperature~\cite{bird07}.
In what follows, the convection contribution was calculated in the approximation of capillary flow at the Nusselt number  $\mathrm{Nu}=2$.
All heat fluxes calculated for Al-12Si and CL20ES powders are provided in Fig.~\ref{fig:heat_fluxes} accounting for actual thermal gradients achieved in SLM processing.

As follows from the result of simulation, the radiative heat flux is by 4 orders of magnitude smaller for Al-12Si  and by 3 orders smaller for the stainless steel than the conductive flux between particles. The obtained conclusions differs with the results in literature~\cite{gusarov10,gusarov09} where radiation is considered as an important physical phenomenon for SLM processing.
However, for partially melted powders the conductive mechanism  substantially exceeds other heat transport mechanisms in the powder bed.

\subsection{Dependence of the thermal conductivity on porosity}
\label{deppor}
\begin{figure}
\begin{center}
\includegraphics[width=0.49\columnwidth]{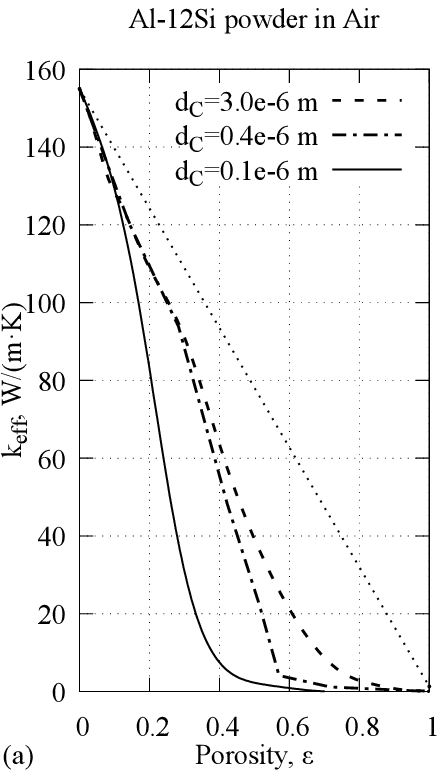}
\includegraphics[width=0.49\columnwidth]{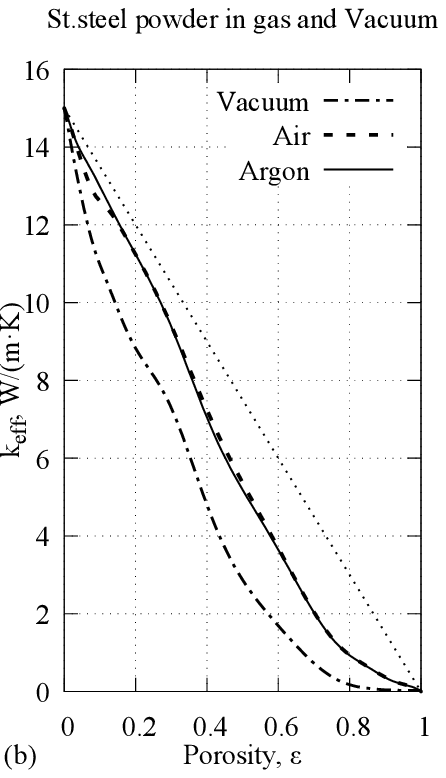}
\end{center}
\caption{\label{f4}  Dependence of the effective thermal conductivity $k_{eff}$ on porosity $\varepsilon$ calculated for (a) Al-12Si powder in air at $d_C=3$~$\upmu$m (dashed curve), $0.4$~$\upmu$m (dash-dotted curve) and $0.1$~$\upmu$m (solid curve); (b) stainless steel CL20ES powder in vacuum (dash-dotted curve), air (dashed curve) and argon (solid curve). This curves were interpolated based on 30 runs with 3 different random-generated  particle configurations for each porosity value. The numerically  calculated points were  smoothed by splines. The dotted lines represent the linear model. }
\end{figure}

Considering a weak dependence of the effective heat conductivity on convective and radiative transfer mechanisms (see Sec.~\ref{conv}), the study was focused on the geometrical conductance of the powder. During consolidation of particles in presence of high temperatures,  metallic powders reduce their porosity via continuous shrinking. Then the contact spot diameter increases. Therefore, assessment for powder layers with different contacts areas in a wide range of porosities $\varepsilon$ is highly demanded. At the moderate values of $\varepsilon$, the generated powder morphology corresponds to  industrial powders (see Sec.~\ref{gen}). The porosity $\varepsilon$ is upper-bounded by $ \varepsilon < 0.26$ which corresponds to dense packing where drastical improvement of the contact surface between particles is expected. There is also a bottom bound at $ \varepsilon > 0.7$ which corresponds to the percolation limit \emph{i.e.} to a break of continuous heat transfer through the metallic phase.

The dependence of the effective thermal conductivity of Al-12Si powder on the porosity $\varepsilon$ for different contact spot diameters $d_C$ calculated in air is presented in Fig.~\ref{f4}(a). Small contact spots (see the solid curve at $d_C=0.1$ $\upmu$m) correspond to the case of dry-powders where a relative size of the contact spots is about $d_C/\langle d \rangle = 0.02$. As follows from the plot, the curve sharply tends to very small values in the range $k_{eff}\simeq 2.5..3.2$ W/(m $\cdot$ K) at $\varepsilon > 0.4$ which represents a non-consolidated powder bed before laser SLM processing. Such low thermal conductivity provides  poor heat dissipation and leads to heat accumulation in the uppermost surface layer. A slight increase of the contact diameter between particles (see curves $d_C=0.1$  and $d_C=0.4$ $\upmu$m in Fig.~\ref{f4}(a)) leads to increase of the heat flux between particles by 16 times.

In general, the dependence of $k_{eff}$ on $\varepsilon$ has almost linear behaviour for porosities smaller than the percolation threshold. Due to random distribution of particles in the bed, heat conduction is reduced to only gas-phase conduction earlier than the percolation limit is achieved. In case of highly consolidated powder (see dashed curve $d_C=3.0$ $\upmu$m in Fig.~\ref{f4}(a)), one gets a curve abutting to the theoretical percolation threshold which is close to the solution for contacted soft spheres (see \cite{Xu2018} and references therein).

Heat conduction in partially consolidated powder for $d_C \geqslant 0.4 \upmu$ ($d_C/\langle d \rangle>0.08$) which correspond to high-density packed powder (briquette) at $\varepsilon<0.3$ is close to the geometrical limit of heat conductivity (Fig.~\ref{f4}(a)). Dry powder at $d_C/\langle d \rangle \leqslant 0.02$ reaches such high conductance only at $\varepsilon<0.2$. The inflexion point in the curve at $\varepsilon \simeq 0.25$ matches to the HCP packing limit.

The effect of different gaseous atmospheres is presented in Fig.~\ref{f4}(b). First, in powders with relatively small porosity or with relatively high bulk thermal conductivity the presence of air or protective atmosphere in pores is insignificant. In calculations with
the highly conductive Al-12Si powder, the influence of gas atmosphere on heat conductivity was found negligible. Therefore, the calculated curves $k(\varepsilon)$ for argon and air coincided. Second, in a case of powders with low conductivity like stainless steel this effect becomes valuable and a significant difference between $k(\varepsilon)$ curves follows from computations (see Fig.~\ref{f4}(b)). The $k_{eff}(\varepsilon)$ curve in vacuum differs from ones for argon and air. All curves show the decay abutting to the theoretical percolation threshold as for Al-based alloys.

Thus the following conclusions can be drawn. Influence of gas  atmosphere on thermal conductivity cannot be neglected for $k^\textrm{metal}/k^\textrm{gas}<1000$. This outcome corresponds to the previous studies~\cite{kaviany,hadley86} and theoretical models as confirmed in what follows in Fig.~\ref{img:porous-comparison}. The effective heat conductivity of a powder bed processed in argon and air are very close owing to similar thermal diffusivities of the gas atmospheres. From a practical standpoint, the calculated $k_{eff}(\varepsilon)$ functions allow one to quantify the effect of dynamically changed porosity and the consolidation coefficient during powder compaction in SLM/SLM processing.

\subsection{Dependence on the consolidation coefficient}

\begin{figure}[h]
  \center
\includegraphics [width=\columnwidth] {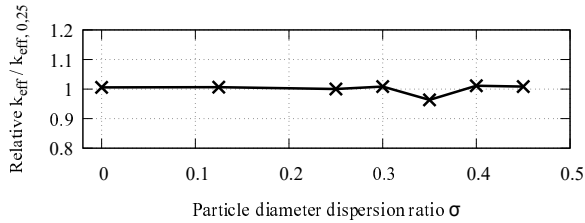}
  \caption{Relative effective thermal conductivity~$k_{eff}/k_{eff,0.25}$ calculated at  different variance~$\sigma$ of particle sizes  in the log-normal size distribution with the mean diameter~$\langle d \rangle$. The randomly generated configurations of particles with a fixed porosity~$\varepsilon \approx 0.4$ was used. The plot is scaled to~$k_{eff,0.25}$ which is the thermal conductivity at~$\sigma=0.25$. The dimensionless overlapping coefficient was kept constant as~$\xi=0.125$ in all cases.}
  \label{img:dispersnost-powder}
\end{figure}

The effect of random distribution of particles on effective thermal conductivity of Al-12Si powder with the average particle size $\langle d \rangle=4.9$~$\upmu$m in presented in Fig.~\ref{img:dispersnost-powder}. The plot shows that this effect is small if the average porosity $\varepsilon$ is fixed.
Then the variation of particle sizes also results in variation of the diameter of the contact area between neighbor spheres.
Thus the small value of variance~$\sigma$ used in the vast majority of calculations is a reasonably representative. In principle, a high value of variance~$\sigma$ in the size distribution allows one to increase the powder packing density and hence its resulted effective thermal conductivity. For instance, the decrease in average porosity from~$\varepsilon=0.55$ to~$\varepsilon=0.40$ allows to double the effective conductivity (see Fig.~\ref{f4}(b) for vacuum). Considering the results of Sec.~\ref{deppor} one can assume that the basic limiting factors for effective heat transfer in powders are both  porosity and the contact spot diameter. Increase of the size variance~$\sigma$ helps to reach denser packing which facilitates heat flow in porous structures.

\begin{figure}
  \center
\includegraphics [width=\columnwidth] {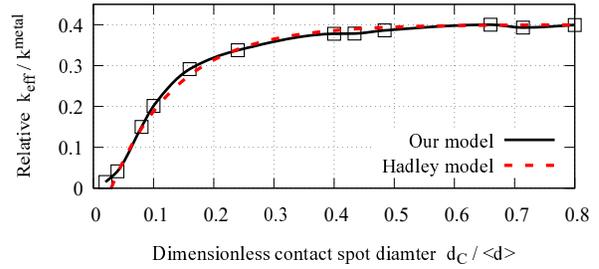}
  \caption{Dependence of the relative thermal conductivity $k_{eff}/k^\textrm{metal}$ on the relative contact spot diameter $d_C / \langle d \rangle$ (see scheme in Fig.~\ref{f1}) calculated for Al-12Si in air. Here $k^\textrm{metal}$ corresponds to the conductivity of bulk metal. The rectangle points and solid black curve denote the results obtained from direct numerical simulations at $\varepsilon \approx 0.55$, red dashed curve corresponds to the Hadley model Eq.~(\ref{eq:Hadley}) with substituted interpolation Eq.~(\ref{alph}) of consolidation coefficient (see Sec.~\ref{exist}). }
  \label{img:peresheek}
\end{figure}

Influence of the contact area between particles with a contact spot diameter $d_C$ on an effective thermal conductivity is analyzed in Figure~\ref{img:peresheek}. In accordance to the analytical model of the elementary cell by Kaviany~\cite{kaviany}, under conditions of week overlapping with~$d_C/\langle d \rangle < 0.1$ a significant descent of thermal conductivity of a metal porous structure occurs. Further decrease of~$\xi$ leads to full regression of the contact area and a transition to the dry powder limit of non-consolidated particles occurs. At moderate $d_C/\langle d \rangle$, a weakly increasing   function of the dimensionless thermal conductivity on the overlapping coefficient is observed in air. In vacuum, there is no conductive  heat transport hence the geometrical effect of the contact zone is even more emphasized. The range of the overlapping coefficient $\xi > 0.2$ geometrically conforms with the preheated powder where neck formation has been already completed or with a case of rough consolidation during SLS and SLM processing on later stages of powder compaction. 
We shall especially note that for the same values of $d_C$ the thermal conductivity could differ significantly owing to the different contact resistance and may be a subject of dynamic change in the SLM process. For example, the analytical assessment of  \cite{kaviany,Chan1973} gives the value of the contact resistance around $\approx 1/(k^\textrm{metal} d_C)$ for sphere-to-sphere Hertzian contact in vacuum.  The overall effect on the thermal conductance with surface effects such as oxide film for the above average contact pressure may be assessed as  $\approx 0.8\times10^{-4}$ W m$^{-2}$ K$^{-1}$ for aluminum alloys and $\approx 0.3\times10^{-4}$ W m$^{-2}$ K$^{-1}$ for stainless steels \cite{Stewart1973,Yovanovich2005}. These effects are very important in the region of small contact areas $d_C$'s for dry powders and shall be disclosed in detail in future study in connection to SLM utilized powders.

In laboratory experiments~\cite{shishkovskiy16,Anestiev1999} on sintered partially-melted powders, heat transport occurs under conditions of randomly distributed particles where~$d_{\textrm{C}}/\langle d \rangle=0.1..0.3$. In the present work, the pre-processed Al-12Si powder with similar characteristics was experimentally tested to visually assess the rate of solid-state consolidation, see Fig.~\ref{fig1}(b). In SLM processing, significant consolidation is provided by either preheating of powder or by its plastic deformation during leveling of the powder layer. At temperatures close to the melting temperature~$T_m$ the particles rapidly start to conglomerate. At the beginning, the total flux through the metal matrix is sufficient to provide initial sintering between particles. As a result, the role of specific contact zones quickly decreases thanks to an increasing number of conglomerated particles. Thus at the late stages of sintering of the metal matrix with low porosity~$\varepsilon$ the effect of the overlapping coefficient and the area of the contact zone decays. To conclude, according to the present calculations at the moderate porosities $\varepsilon=0.4..0.6$ the influence of the contact zone area for a preheated sintered powder for $d_C/\langle d \rangle \geqslant 0.2$ is little. In the range $d_C/\langle d \rangle < 0.2$, thermal conductivity of the powder drastically drop to zero as the contact zone becomes smaller.

\subsection{Comparison to existed models}
\label{exist}

\begin{figure}[!h]
  \center
\includegraphics [width=\columnwidth] {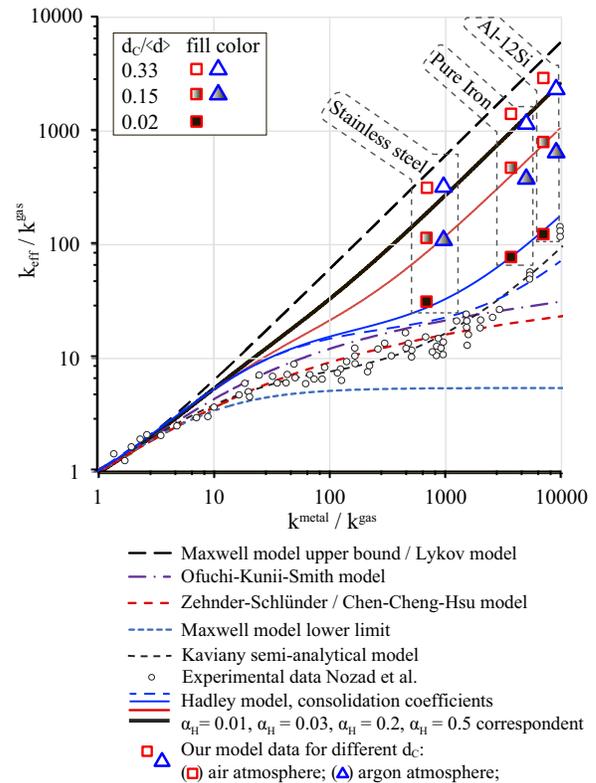}
\caption{Comparison between the analytical models of effective heat conductivity, experimental data by Nozad et al.\cite{nozad1, nozad2} and numerical simulations carried out in the present work at $\varepsilon \approx 0.38$ for powders with spherical particles. The curves depicted in the figure are the models of Maxwell upper bound (Lykov model), Ofuchi-Kunii-Smith, Zehnder-Schl{\"{u}}nder, Chen-Cheng-Hsu, Maxwell lower bound, Kaviany and Hadley \cite{likov67, kunii60, ofuchi64, zehnder70, hsu94, kaviany, hadley86} correspondingly. The present results are depicted by the squares and triangles (for different gas atmospheres) in different dashed boxes for stainless steel, iron and Al-12Si. The shape and color of the symbols correspond to different contact spot diameters $d_C/\langle d \rangle$ and gas atmosphere used in simulations. The figure is based on the diagram by Henriksen~\cite{henriksen13}.}
  \label{img:porous-comparison}
\end{figure}

Comparison of our numerical simulations with different analytical models is provided in Fig.~\ref{img:porous-comparison}. The plotted diagram gives the ratio $k_{eff}/k^\textrm{gas}$ of the effective thermal conductivity scaled to the thermal conductivity of gas in the $y$-axis. The ratio $k^\textrm{metal}/k^\textrm{gas}$ of the thermal conductivities of metal and gas is plotted in the $x$-axis. The diagram is depicted in the log-log scale and it was calculated at the fixed porosity $\varepsilon=0.38$.

The Lykov model also known in literature as the Maxwell upper bound model for thermal conductivity of two-phase media~\cite{kaviany,likov67} yields the upper geometrical limit of conductivity. The Maxwell lower bound corresponds to the lower limit of $k_{eff}/k^\textrm{gas}$  where heat transfer is mainly controlled by thermal conductivity through the gas phase.

The models of  Zehnder-Schl{\"{u}}nder \cite{zehnder70} and Chen et al. \cite{hsu94} are very close to each other for the given $\varepsilon$ and are plotted by the same line. These models are of the elementary-cell type (also the Ofuchi-Kunii-Smith~\cite{kunii60,ofuchi64} model) and consider heat transfer in powders in the approximation of point contacts between particles. This approach is valid for dry-powders and it results in the  $k_{eff}$-curves  located lower than our data for partially-melted powder. The model of Kaviany \cite{kaviany} considers finite necks between powder particles and it allows to describe high-conductive powders. It is worth mentioning that the semi-empirical Kaviany model is conceptually close to our approach because both models account for formation of necks with variable cross-section areas between particles.

The Hadley model suggested in~\cite{hadley86} introduces a consolidation coefficient $\alpha_H$ which in fact is a conversion coefficient to match two limiting cases of the Maxwell upper and lower bound models:
\begin{align}\label{eq:Hadley}
\frac{k_{eff}}{k^\textrm{gas}}=  \left( 1-\alpha_H  \right)\! {\frac {   \varepsilon{\it f_0}+ \left( 1-\varepsilon{\it f_0} \right) \frac{k^\textrm{metal}}{k^\textrm{gas}}  }{1-\varepsilon \left( 1-{\it f_0} \right) +\frac{k^\textrm{metal}}{k^\textrm{gas}}\varepsilon \left( 1-{\it f_0} \right) }}+ \nonumber  \\
{+ \alpha_H \frac {  2 \left( 1-\varepsilon \right) \left( {\frac{k^\textrm{metal}}{k^\textrm{gas}}}\right)^{2}  + \left( 1 +2\,\varepsilon \right) \frac{k^\textrm{metal}}{k^\textrm{gas}}  }{1-\varepsilon +  \left( 2+\varepsilon \right) \frac{k^\textrm{metal}}{k^\textrm{gas}}}} .
\end{align}
where ${\it f_0}$ is an empirical constant, which is selected as ${\it f_0}=0.9$ in all our cases. This model allows one to describe both partially melted and dry-powders in a single coherent approach. At $\alpha_H=0.01$, the media is loose-packed and the curve plotted in Fig.~\ref{img:porous-comparison} is very close to the data of Nozad et al.~\cite{nozad1,nozad2} for dry powders. Our computations are presented in Fig.~\ref{img:porous-comparison} by red squares and blue triangles in the columns combined for metallic powders of stainless steel, Al-12Si and pure iron. The parameters of iron were taken from~\cite{desai86,crc} and the mean diameter was $\langle d \rangle = 5$~$\upmu$m.

The results of present calculations in a case of partially-melted particles conform well to the Hadley model at intermediate and large $\alpha_H \geqslant 0.03$.
Further, in consolidated powders with good thermal contact between particles the dependence of the effective thermal conductivity $k_{eff}$ on the contact spot diameter $d_C$ is sufficiently small (see Fig.~\ref{img:peresheek} in the interval $d_C/\langle d \rangle \geqslant 0.2$). This conclusion has been also confirmed by both the Hadley's and our models.
 For smaller $\alpha_H<0.03$, the function $k_{eff}/k^\textrm{gas}$ depends on the contact spot diameter non-linearly. The contact area decreases as square of the contact spot diameter and hence the heat flux between the particles drastically deteriorates as was shown in Fig.~\ref{img:peresheek} in the interval $d_C/\langle d \rangle < 0.2$. To link our model to the analytical expression we propose a simple approximation of the consolidation coefficient $\alpha_H$:
 \begin{equation}
\label{alph}
\alpha_H = 1-\exp \left(-A \left( {d_C}/{\langle d \rangle } - d_0 \right) \right) ,
\end{equation}
where $A$ is pre-exponential factor and $d_0$ is a dry-powder shift. In Figure~\ref{img:peresheek}, one can find fair agreement between the numerical data (black curve) and the resulted function $k_{eff}/k^\textrm{metal}$ analytically calculated using Eq.~(\ref{eq:Hadley}) with substituted Eq.~(\ref{alph}) at $A=9$, $d_0=0.03$, $\varepsilon=0.5$ (red dashed curve).

The smooth rise of $k_{eff}$ in Fig.~\ref{img:porous-comparison} reported by data of Nozad et al., Kaviany and Hadley models for highly heat-conductive metals such as pure iron and Al-based alloys has also been reproduced in our calculations.  The black rectangular points in Fig.~\ref{img:peresheek} correspond to weak thermal contact between particles at $d_C/\langle d \rangle \simeq 0.02$. There is a clear point of inflection in the $k_{eff}$ curves of the Hadley model. The inflection occurs at about $k^\textrm{metal}/k^\textrm{gas} \sim 1000$.
This transition in the $k_{eff}$ is not described well by the elementary-cell models.

One can conclude that good agreement with the Hadley model has been received in our simulations for consolidated powders. In the dry-powder limit, correspondence is worse. Therefore, we think that the numerical model has better accuracy in the account of local morphology and the presented approximation of the consolidation coefficient will lead to more robust analysis in wide interval of parameters of partially consolidated powders.

\section{Conclusions}

The effective thermal conductivity $k_{eff}$ as a function of porosity $\varepsilon$ for Al-12Si and stainless steel powders was calculated in a series of computer simulation runs. The obtained results account for the effects of consolidation which can drastically change heat conductivity of powders in SLS/SLM processing.

The obtained effective heat conductivities for commercial powders are suitable for rapid  modeling of powder consolidation and optimization of selective laser
sintering/melting. The temperature dependence of $k_{eff}$  can be introduced by multiplication of the normalized $k_{eff}/k^\textrm{metal}$ function on the bulk heat conduction coefficient which is dependent on temperature.

It was found that the function $k_{eff}(\varepsilon)$ is a nonlinear curve which has three distinctive intervals. At porosity $\varepsilon < 0.26$, the powder layer has dense packing (briquette) where the contact spots between particles are developed and drastically
improvement of heat conductivity is observed. The calculated $k_{eff}(\varepsilon)$ curves have quasi-linear behavior $k_{eff} \sim k_0 (1-\varepsilon)$, Fig.~\ref{f4}. In the intermediate interval of porosities, strong dependence on the contact spot diameter $d_C$ is registered. Hence, the $k_{eff}(\varepsilon)$ curves are shifted to lower values comparing to the linear model.
Also there is a bottom bound at $ \varepsilon > 0.7$ in Fig.~\ref{f4}. which corresponds to the
percolation limit \emph{i.e.} where a break of continuous heat transfer through the metallic phase occurs.

The effect of gas atmosphere is as follows. The thermal conductivity of stainless steel powder is affected by heat transfer through the gas phase. At the same time, Al-12Si powder is less
sensitive to heat conduction through gas. In the paper, the criterion for accounting of the heat conductance in gas has been suggested as $k^\textrm{metal}/k^\textrm{gas} < 10^3$. There is no substantial difference in $k_{eff}$ between the powders processed in argon and air.

The effective heat conductivity of partially consolidated particles steeply increases if the dimensionless diameter of thermal contacts becomes $d_C/\langle d \rangle>0.2$, Fig.~\ref{img:peresheek}. The heat flux through a contact spot depends on its diameter $d_C$ as a quadratic function, thus in the range $0.2 < d_C/\langle d \rangle < 0.8$  a weak dependence of $k_{eff}$ on $d_C$ is obtained.  The nonlinear dependence is found at $d_C/\langle d \rangle < 0.1$ where the degree of consolidation becomes smaller  approaching the dry-powder limit of thermal conductivity.

Based on the performed analysis, we suggested a new analytical fitting Eq.~(\ref{alph}) which links the consolidation coefficient $\alpha_H$ of the Hadley model~\cite{hadley86} Eq.~(\ref{eq:Hadley}) to the realistic powder parameters which are the diameter of contact necks $d_C$ and mean particle diameter $\langle d \rangle$.
The results of numerical simulations show that the distribution of particle in size does not affect $k_{eff}$ explicitly. The increase of the distribution variance leads to a more dense packing with a smaller porosity.

Suggested coefficients $k_{eff}$ were derived to use in the approximation of continuous media. Thus, it is valid only under the conditions of local thermal equilibrium. For SLS/SLM processes, the characteristic scale of applicability of $k_{eff}$ is limited by $L>200$ $\mu$m for macroscopic models due to high temperature gradients observed in the powder layer.

In summary, the suggested numerical model broadens the description of transport characteristics of partially melted powders. It substantially extends the interval of applicability of the analytical models from dry-powders to fully-consolidated powders. This allows to evaluate more precisely heat transfer in the powder layer during its  compaction in SLM processing. Also the thermal
conductivity of preheated sintered powder can be assessed.

\begin{acknowledgment}
This study was  financially supported by Russian Science Foundation, project 19-79-20012.
\end{acknowledgment}

%
% ...

\bibliographystyle{asmems4}

\bibliography{mybibfile}



\end{document}